\documentclass[12pt]{article}
\usepackage{erice,epsfig}

\begin{document}
\begin{center}
\textbf{``On the Theory of Collisions between \\ Atoms and Electrically Charged 
Particles''}
\end{center}

\begin{center}
\textbf{Note by: Enrico Fermi, 1924}
\end{center}

\begin{center}
\textbf{Appeared in Nuovo Cimento 2, pp. 143-158 (1925)}
\end{center}

\begin{center}
\textbf{(translated from Italian
by Michele Gallinaro and Sebastian White,
New~York 2001)}
\end{center}

\vspace*{2cm}

\noindent
\textbf{Translators' Note:}
\vspace*{0.3cm}\\
In the fall of 1924, Enrico Fermi visited Paul Ehrenfest at Leyden on a 3-month
fellowship from the International Education Board (IEB). Fermi was 23~years old.
In his trip report to the IEB, Fermi says he learned a lot about cryogenics and
worked on two scientific papers, including the following one. It was submitted
in German to Zeitschrift f\"ur Physik. The German version was known to Weizs\"acker
and Williams and cited in the papers (10 years) later in which they extended
Fermi's method to the Ultra-Relativistic case. The German version was
subsequently translated into a Russian version and perhaps other
languages. Fermi's Italian version (printed in {\it Nuovo Cimento}) is less
widely known and does not appear in the ``Collected Works''. Nevertheless,
Persico remarks that this was one of Fermi's favorite ideas and that he often
used it in later life. So, we would like to think of this as a late 100$^{th}$
birthday present to the Italian Navigator.\\
We would like to thank Professor T.D.~Lee for his encouragement of this
project and for interesting discussions about
Fermi. Also Tom Rosenblum at the Rockefeller
Archives for bringing Fermi's correspondence to our attention and Bonnie
Sherwood for typing the original manuscript.

\newpage

\begin{center}
\textbf{``On the Theory of Collisions between \\ Atoms and Electrically Charged 
Particles''}
\end{center}

\begin{center}
\textbf{Note by: Enrico Fermi, 1924}
\end{center}

\vspace*{2cm}

I. When an atom, in its normal state, is illuminated with light of the 
appropriate frequency, it can be excited, which is to say that it goes to a 
quantum state of higher energy, absorbing a quantum of light. If the quantum 
of the exciting light is greater than the energy needed to ionize the atom, 
it can be ionized losing, depending on the light frequency, an electron of 
either the superficial layers or those deep within the atom. During 
the phase of relaxation of the atom there is emission of light and, 
depending on the circumstances, it is possible to obtain the phenomena of 
optical resonance or fluorescence. Some natural phenomena, very similar to 
these, occur also during excitation by collision.

In fact, if the atoms in a gas are bombarded with electrons of high enough 
velocity, they can be excited or ionized. They can also lose, if the 
velocity of the bombarding electrons is very large, some electrons belonging 
to the innermost layers of the atom.

The scope of the present work is to further specify the analogies (existing) 
between these two classes of phenomena, and to quantitatively derive the 
phenomena of excitation by collision from those of optical absorption. 
Therefore, we consider that when a charged particle passes near a point, it 
produces, at that point, a variable electric field. If we decompose this 
field, via a Fourier transform, into its harmonic components we find that it 
is equivalent to the electric field at the same point if it were struck by 
light with an appropriate continuous distribution of frequencies. If we now 
imagine that, at that point, there is an atom, the hypothesis occurs quite 
naturally, that the electric field of the particle produces on the atom 
those same effects of excitation or ionization that the equivalent light 
would produce.
Let us suppose that we know the absorption coefficient for light by the 
atom, as a function of frequency. We would then have a way to calculate the 
probability that a charged particle, passing with a given velocity at a 
given distance from an atom, will ionize it. It seems, on the other hand 
absolutely necessary to introduce a limitation to this correspondence 
between the electric field of the light and that of a charged particle. We 
note, in fact, that a particle of velocity, $v$, cannot produce phenomena, by 
collision, which require an energy greater than its own kinetic energy. If, 
instead, we perform a harmonic decomposition of the electric field it 
produces, we see that it will include all possible frequencies, including 
the highest ones. Therefore we must accept that all frequencies whose 
quantum, $h\nu $, is greater than the particle's kinetic energy can have no 
effect (since not enough energy is available to supply an entire quantum).

We have applied our hypothesis to three phenomena that allow experimental 
verification. They are:

\noindent
a) Excitation of the line 2537 (a resonance line) of mercury. The optical 
absorption of this line is known with sufficient precision. Experiments also 
exist that give, at least to an order of magnitude, the probability that 
atoms of mercury bombarded with slow electrons are excited. Our theory gives 
for this probability the right order of magnitude. Naturally, given the 
limited precision of the experimental data, a precise test of the theory 
through these phenomena is not possible, but the case nonetheless is of some 
interest since no other theory of the excitation, by collision, of a quantum 
state exists, only theories of ionization by collision.

\noindent
b) The number of ion pairs produced per cm of path length by $\alpha 
$-particles from Radium-C in Helium. We chose Helium because, being a 
monoatomic gas, and possessing only k-shell electrons, it's possible to know 
with some precision its absorption coefficient as a function of frequency. 
The agreement between experiment and theory is very good. The previous 
theories of ionization by 
collision~\footnote{J.J. Thomson - Phil. Mag. 23, p.~449, 1912.

	  ~~~N. Bohr - Phil. Mag. 30, p.~581, 1915.

	  ~~~S. Rosseland - Phil. Mag. 45, p.~65, 1923.}
are all based on a principle different 
from the present theory. The essential point of those theories can be 
summarized in the following way: when a charged particle passes within the 
vicinity of an atom it transfers to the electrons of the atom a fraction of 
the particle's energy. The energy transfer is calculated assuming, 
typically, that the electrons are free within the atom. One then allows the 
electron to be detached from the atom every time that the energy transferred 
to it via this mechanism is greater than the ionization energy of the atom.

\noindent
c) The path length in Helium of $\alpha $'s from Radium-C. The mechanism of 
this phenomenon is, according to our theory, the following: the $\alpha $ 
particle creates, along its path, a variable electric field whose energy is 
absorbed by the surrounding atoms, as if it were the electric field of a 
light wave. The energy, which is thus absorbed, has to obviously be 
subtracted from the kinetic energy of the $\alpha {\rm t}$particle which 
is thus decelerated. Also, in this case, the agreement between the 
theoretical and experimental values is excellent. Previously, the 
deceleration of $\alpha $ particles in matter had been studied theoretically 
by Thomson and by 
Bohr~\footnote{N. Bohr - Phil. Mag. 25, p.~10; 1913, Phil. Mag. 30, p.581; 1915.}. 
The former interpreted it as due to the fact that 
the $\alpha $ particle, attracting electrons of the surrounding atoms along 
its path, transfers to them a fraction of its kinetic energy. Since, on the 
other hand, when one calculates this energy as if the electrons were free, 
one finds an infinitely large deceleration, Bohr redid the calculation 
assuming that the electrons inside the atoms were no longer free, but bound 
by quasi-elastic forces. In this way he finds a good agreement with 
experiment, taking for the frequency of this binding, as the case may be, 
optical or {\it x}-ray (R\"ontgen) frequencies.

In the second section of this work we establish some general formulas for 
the calculation of the ``field of light'' that produces an electric field 
equivalent to that of the particle.
In the third, fourth and fifth sections we discuss the application to a), b) 
and c).

II. Let's calculate, first of all, the spectral distributions corresponding 
to those of the electric field created by a particle with electric charge, 
$\varepsilon $, passing with velocity, $v$, at a minimum distance, $b$, from a 
point, $P$. The components of the electric force, at $P$, parallel and 
perpendicular to the motion of the particle, are

\begin{equation}
 E_1 = \frac{\varepsilon v\theta }{(b^2 + v^2\theta ^2)^{\frac{3}{2}}},
\quad E_2 = \frac{\varepsilon \cdot b}{(b^2 + v^2\theta ^2)^{\frac{3}{2}}}~~.
\end{equation}

Where $\theta $ is the time-elapsed starting from the passage of the 
particle at the minimum distance from $P$. Taking now $T$ to be a time that we 
will later extend to infinity, we can expand $E_{1}$ and $E_{2}$ in a 
Fourier series for all values of $\theta $ between --(1/2)$T$ and +(1/2)$T$.\\
We then find

\begin{equation}
 E_1 = \sum {a_n } \sin \frac{2\pi n}{T}\theta  ,\quad E_2 = \sum {b_n } 
\cos \frac{2\pi n}{T}\theta 
\end{equation}

\noindent where

\begin{equation}
a_n = \frac{2\varepsilon v}{T}\int_{ - \frac{T}{2}}^{\frac{T}{2}} 
{\frac{\theta \sin \frac{2\pi n\theta }{T}d\theta }
{(b^2 + v^2\theta ^2)^{\frac{3}{2}}}},\quad b_n = \frac{2\varepsilon b}{T}
\int_{ - \frac{T}{2}}^{\frac{T}{2}} {\frac{\cos \frac{2\pi n\theta }{T}d\theta }{(b^2 
+ v^2\theta ^2)^{\frac{3}{2}}}} ~~.
\end{equation}

The electric field, $a_{n}$~sin(2$\pi n\theta /T$) is now equal to the 
electric field of a light wave with intensity $(c/4\pi)\cdot (a^2_n/2)$
and frequency $\nu =n/T$. We now let $I(\nu )d\nu $ be the light 
intensity in a frequency interval d$\nu $, multiplied by the duration, $T$, of 
its action. We find

\[ I\left( \nu \right)d\nu = 
 {{\lim} \atop {T \to \infty }} ~
\frac{cT}{4\pi }\frac{a_n^2 + b_n^2 }{2}Td\nu = 
\]

\[
 = \frac{c}{2}\frac{\varepsilon ^2}{\pi }\left\{ {b^2\left( {\int\limits_{ - 
\infty }^\infty {\frac{\cos 2\pi \nu \theta d\theta }{\left( {b^2 + 
v^2\theta ^2} \right)^{\frac{3}{2}}}} } \right)^2 + v^2\left( {\int\limits_{ 
- \infty }^\infty {\frac{\theta \sin 2\pi \nu \theta d\theta }{\left( {b^2 + 
v^2\theta ^2} \right)^{\frac{3}{2}}}} } \right)^2} \right\}d\nu .
\]

\noindent
The integrals that appear in this expression can be represented through the 
modified Bessel 
functions~\footnote{See, for example, Gray, Mathew and Macrobert - 
Bessel Functions - London, 1922, where also numerical tables of the values of
these functions can be found. For the Fourier series expansions of these
expressions, see also Bohr, l.e.} 
of the second kind of order zero and one. Explicitly we have

\[
\int\limits_{ - \infty }^\infty {\frac{\cos \omega xdx}{\left( {1 + x^2} 
\right)^{\frac{3}{2}}}} = 2\omega K_1 \left( \omega \right)
\]

\[
\int\limits_{ - \infty }^\infty {\frac{x\sin \omega xdx}{\left( {1 + x^2} 
\right)^{\frac{3}{2}}}} = - 2\omega K_0 \left( \omega \right) .
\]

With these formulas we immediately find

\begin{equation}
I\left( \nu \right) = \frac{8\pi c\varepsilon ^2\nu ^2}{v^4}\left\{ 
{K_0 ^2\left( {\frac{2\pi \nu b}{v}} \right) + K_1 ^2\left( {\frac{2\pi \nu 
b}{v}} \right)} \right\}
\end{equation}

\noindent where we substitute for brevity

\[
B\left( \omega \right) = K_0 ^2\left( \omega \right) + K_1 ^2\left( \omega 
\right)
\]

\noindent
and we then find

\begin{equation}
I\left( \nu \right) = \frac{8\pi c\varepsilon ^2\nu ^2}{v ^4}B\left( 
{\frac{2\pi \nu b}{v }} \right) .
\end{equation}

Since all the frequencies whose quantum is larger than the kinetic energy of 
the particle cannot, according to our hypothesis, have any effect, we find 
finally that the charged particle's passage is equivalent to the following 
spectral distribution

\begin{equation}
I\left( \nu \right) = \left\{ {{\begin{array}{*{20}c}
 {\frac{8\pi c\varepsilon ^2\nu ^2}{v ^4}B\left( {\frac{2\pi \nu b}{v }} 
\right),\quad h\nu \le \frac{1}{2}mv^2} \hfill \\
 {0 \qquad \qquad \qquad  ,\quad h\nu > \frac{1}{2}mv^2} \hfill \\
\end{array} }} \right.
\end{equation}

III. We now want to show, using a simple example, the application of the 
general method: therefore we will study the excitation by collision of 
spectral lines. For example, we consider an atom of mercury, and take 
$\alpha (\nu )$ to be its absorption coefficient at the frequency $\nu $. 
This means that if the atom is struck by light of intensity $I(\nu )d\nu $ 
it absorbs, on average, the energy $\alpha (\nu )I(\nu )d\nu $ 
and therefore the absorption probability of a quantum $h\nu $ is

\begin{equation}
\frac{\alpha \left( \nu \right)I\left( \nu \right)d\nu }{h\nu } ~~.
\end{equation}

Since the absorption at the resonance line corresponds to the excitation of 
the atom to the 2$p$ state, we find that (7) represents the excitation 
probability if only the frequencies in the interval $d\nu $ would be 
effective. The probability $\Pi (b)$ for the excitation to be caused by the 
passage of the particle at a distance, $b$, is naturally smaller than the sum 
of all the probabilities obtained from the individual frequency intervals 
$d\nu $ when one treats these as independent. 
One easily finds that~\footnote{Effectively the probability that the atom
doesn't absorb any quantum is
\[
1 - \Pi \left( b \right) = \prod {\left( {1 - \frac{I\left( \nu 
\right)\alpha \left( \nu \right)d\nu }{h\nu }} \right)} 
\]
in which the product must be extended over all the infinitesimal intervals 
$d\nu $.

Taking the logarithms of both sides we find in the limit 
\[
\log \left( {1 - \Pi \left( b \right)} \right) = \sum {\log \left( {1 - 
\frac{I\left( \nu \right)\alpha \left( \nu \right)d\nu }{h\nu }} \right)} = 
- \int { \frac{I\left( \nu \right)\alpha \left( \nu \right)d\nu }{h\nu }} 
\]
from which we obtain Eq.~(8).}

\begin{equation}
\Pi ( b ) = 1 - e^{ - \int {\frac{I( \nu )\alpha ( \nu )d\nu}
{h\nu }}}.
\end{equation}

Since, in the case of the resonance, $\alpha (\nu )$ has a value different 
from zero only in a very narrow band near the resonance frequency $\nu _{0}$, 
we can write 

\[
\int {\frac{I\left( \nu \right)\alpha \left( \nu \right)d\nu }{h\nu }} = 
\frac{\alpha }{h\nu _0 }I\left( {\nu _0 } \right)
\]

\noindent
where we use

\[
\alpha = \int {\alpha \left( \nu \right)d\nu } .
\]

Equation~(8) can therefore be written, in this case as

\[
\qquad\qquad\qquad\qquad\qquad\quad\quad\quad \Pi \left( b \right) = 1 - e^{ - \frac{\alpha }{h\nu _0
}I( {\nu _0 } )} . \qquad\qquad\qquad\qquad\qquad\qquad\quad\quad (8^\prime)
\]

To derive from this an observable directly accessible to experiment, we 
calculate the effective radius, $\rho $, of the atom for
resonant excitation. That is to say the equivalent radius needed to obtain 
the same overall excitation probability if the excitation probability where 
unity within this radius. This radius is evidently given by

\[
\pi \rho ^2 = 2\pi \int b~ db\Pi \left( b \right) .
\]

We therefore find, keeping in 
mind~\footnote{Effectively an electron, which passes with low velocity near an atom, 
will be strongly scattered by it. Since we are only after the order of 
magnitude, we'll apply Eq.~(6) anyway.} Eq.~(6), that for 
$\frac{1}{2}mv^2 > h\nu $

\begin{equation}
\rho ^2 = 2\int\limits_0^\infty {\left\{ 
 {1 - e}^{ - \frac{8\pi \varepsilon ^2c\alpha \nu _0 }{hv^4}B\left( {\frac{2\pi 
\nu _0 b}{v }} \right)}  \right\}} bdb
\end{equation}

\[
 = \frac{v ^2}{2\pi ^2\nu _0 ^2}\int\limits_0^\infty {\left\{ {1 - 
e{\begin{array}{*{20}c}
 { - \frac{8\pi \varepsilon ^2c\alpha \nu _0 }{hv^4}B\left( x \right)} 
\hfill \\
 \hfill \\
\end{array} }} \right\}} xdx .
\]

The integral can be evaluated approximately~\footnote{To find an
approximate expression for

\[
I\left( \alpha \right) = \int\limits_0^\infty {\left( {1 - e{\begin{array}{*{20}c}
 { - \alpha B\left( x \right)} \hfill \\
 \hfill \\
\end{array} }} \right)} xdx
\]

\noindent
for $\alpha \ll 1$, we observe that when $x < 0.4$ one has with sufficient 
accuracy  $B(x)=1/x^{2}$. For $\alpha < 1$ and $x > 0.4$ one can set

$$ 1 - e^{ - \alpha B(x) }=\alpha B(x) ~~. $$

We can then write
\[
I\left( \alpha \right) = \int\limits_0^{0.4} {\left( {1 - e{\begin{array}{*{20}c}
 { - \frac{\alpha }{x^2}} \hfill \\
 \hfill \\
\end{array} }} \right)} xdx + \alpha \int\limits_{0.4}^\infty {B\left(x \right)} x dx
\]
The numerical calculation gives us for the second integral $0.973 \cdot \alpha $.

The first can be calculated easily using the asymptotic expressions for the 
logarithmic integral and one finds that it has the value
\[
 - \frac{\alpha }{2}\log \alpha - 0.705\alpha .
\]
We then obtain finally
\[
I\left( \alpha \right) = \left( {0.268 - \frac{1}{2}\log \alpha } 
\right)\alpha .
\]
} and one finds

\[
\qquad\qquad\qquad\qquad\qquad \rho ^2 = \frac{2\varepsilon ^2c\alpha }{\pi h\nu _0
v^2}\left( {0.54  
- \log \frac{8\pi \varepsilon ^2c\alpha \nu _0 }{hv^4}} \right) .
\qquad\qquad\qquad\qquad\qquad  (9^\prime)
\]

Experimentally the excitation by collision of the resonance line 2537 of 
mercury was studied by 
Miss Sponer~\footnote{ Hertha Sponer, ZS. f. Phys. 7, p.185, 1921.}.

Unfortunately this work gives us only the order of magnitude of the 
excitation probability. In fact, Sponer reached the conclusion that in the 
collisions of electrons having velocity not much greater than 4.9 volts, 
with atoms of mercury, inelastic collisions amount to a few percent of all 
collisions. To calculate $\rho $ using Eq.~(9$^\prime$) we take for $v$ a velocity 
corresponding to a potential of 8 volts; the value of $\alpha $ can be 
obtained from a work of 
F\"uchtbauer~\footnote{F\"ucthbauer - Phys. ZS. 21, pp. 322, 694, 1920.}.
This author finds, in fact, that when a 
mercury atom is illuminated with an amount of light having the spectral 
distribution $I(\nu )$, the absorption probability of a quantum of 
``resonance light'' is $PI(\nu _{0})$ where $P = 8\times 10^{-7}$. 
Evidently $\alpha = Ph \nu_{0}$, and therefore $\alpha = 6 \times 10^{ - 4}$. 
Equation~($9^\prime$) gives us then
$\rho = 0.8\times 10^{-8}$.

This value is considerably larger than that found by Sponer. If, in fact, we 
take the fraction of inelastic collisions to be as large as 9{\%} we still 
find, for $\rho $, the value 
$0.4\times 10^{-8}$ which is about half as large as what we calculated.
On the other hand, it is easy to understand the reason for this discrepancy. 
When, an electron of velocity equivalent to a few volts passes at a distance 
of the order of magnitude of $10^{-8}$cms from an electron, it is already 
strongly scattered in such a way that its distance of closest approach is 
considerably larger than if it had remained on its original trajectory. One 
then understands that the error made in neglecting the deflection leads to 
an overestimate of $\rho $. Since, on the other hand, the uncertainty in the 
measurements excludes anyway a precise test of the theory, and with the 
present hypotheses, we obtained the right order of magnitude for $\rho $, it 
seems superfluous to attempt a more precise calculation which certainly 
would be more complicated.

Ionization phenomena by collision can be explained in a very similar way. It 
is known that all atoms beyond the limit of their principal series exhibit, 
both in absorption and emission, a continuous spectrum, corresponding to the 
transition of the valence electron to the state in which it is ionized and 
furthermore possesses some kinetic energy. The spectrum consists of a 
reasonably broad band. It has a sharp cutoff towards the red due to the 
limit of the principal lines and is blurred towards the violet. If, now, the 
velocity of the colliding electron is such that the frequency $mv^{2}/2h$ 
lies beyond the absorption band, it is still possible, as a rough 
approximation, to apply Eq.~(9$^\prime$) also for the calculation of the 
equivalent radius of the atom for ionization effects; if instead the 
limiting frequency falls within the absorption band the radius will 
naturally be smaller than found with Eq.~(9$^\prime$) and will finally go to zero when 
the frequency $mv^{2}/2h$ coincides with or falls below the limit of the 
principal series. Qualitatively this behavior of $\rho $ as a function of $v$ 
is confirmed by the 
experiment~\footnote{Nettleton - Proc. Nat. Acad. of Sci, 10, p. 140, 1924.}. 
Unfortunately it seems impossible to pursue it quantitatively because precise 
data on the intensity of the continuous absorption spectrum are not available.

IV. Instead, it is considerably simpler to calculate the ionization produced 
by $\alpha $ particles. In fact, $\alpha $-particles, because of their 
considerable mass, are hardly deflected and one can, to a good 
approximation, use Eq.~(6). The empirical results on absorption in the 
region of R\"ontgen rays can be summarized with the following 
formula~\footnote{See, for example, H.A. Kramers - Phil. Mag. 46, p. 863, 1923.}

\begin{equation}
\alpha \left( \nu \right) = \frac{KZ^4}{\nu ^3} + D
\end{equation}

\noindent
where $\alpha (\nu )$ represents the atomic absorption coefficient, $D$ is 
the term of the absorption due to [Compton] scattering, $Z$ is the atomic 
number, and $K$ is a coefficient, which exhibits some discontinuities at the 
edge of the series. For values of $\nu $ larger than the threshold,
$\nu _{0}$, of the $k$-series we have $K=0.6\times 10^{30}$, 
while for $\nu <\nu_{ 0}$, $K= 0.1 \times 10^{30}$. 
From here one derives that the contribution to the 
absorption due to ionization from the $k$-shell, is

\begin{equation}
\alpha \left(\nu \right) = \left\{ {{\begin{array}{*{20}c}
 {0,\quad\quad~\nu < \nu _0 } \hfill \\
 {\frac{HZ^4}{\nu ^3},\quad\nu > \nu _0 } \hfill \\
\end{array} }} \right .
\end{equation}

\noindent
$H$ represents, naturally, the discontinuity of the $K$ coefficient when we 
cross the threshold for the $k$ series. One has therefore

\[
H = 0.5 \times 10^{30}.\]

The probability $\Pi (b)$, for an $\alpha $-particle to ionize the $k$-shell 
can be calculated with Eq.~(8). Since the mass of the $\alpha $-particle is 
very large, we can substitute the upper limit $mv^{2}/2h$, with $\infty$. 
Thus we find

\begin{equation}
\Pi \left( b \right) = 1 - e^{
  - \frac{8\pi c\varepsilon ^2HZ^4}{hv^4}\int\limits_{\nu _0 }^\infty 
{\frac{d\nu }{\nu ^2}} B\left( {\frac{2\pi \nu b}{v}} \right)} .
\end{equation}

\noindent
The equivalent radius can be calculated with the formula

\[
\rho ^2 = 2\int\limits_0^\infty {bdb } \Pi 
\left( b \right) .
\]

We find then

\begin{eqnarray}
\rho ^2 & = & 2\int\limits_0^\infty bdb\left\{ 1 - e^{
  - \frac{8\pi c\varepsilon ^2HZ^4}{h v^4}\int\limits_{\nu _0 }^\infty 
\frac{d\nu }{\nu ^2} B\left( \frac{2\pi \nu b}{v} \right)} 
\right\} \\ 
& = & \frac{v^2}{2\pi ^2\nu _0 ^2}\int\limits_0^\infty \xi d\xi \left\{ 1 - 
e^{- \alpha \xi \int\limits_{\xi}^\infty \frac{du}{u^2} B( u 
)}\right\} \nonumber 
\end{eqnarray}

\noindent
where for brevity one puts 

\begin{equation}
\alpha = \frac{8\pi c\varepsilon ^2HZ^4}{h\nu _0 v^4} ~~.
\end{equation}

The integrals in (13) can be calculated approximately~\footnote{We need to calculate 
\[
I = \int\limits_0^\infty \xi d\xi \left\{ 1 - e^{
  - \alpha \xi \int\limits_\xi ^\infty \frac{du}{u^2} B( u )} 
 \right\} .
\]
When $\alpha  < 0.1$ one can break up $I$ into an integral from 0 to 0.4 and 
another from 0.4 to infinity. The first can be calculated observing that 
for $\xi <0.4$ we can substitute with reasonable approximation, 
\[
\int\limits_\xi ^\infty \frac{du}{u^2}  B\left( u 
\right) = \frac{1}{3\xi ^3} - 0.64
\]
\noindent
and then the asymptotic expressions for the logarithmic integral can be 
applied. The second integral can instead be approximately written as 
\[
\alpha \int\limits_{0.4}^\infty \xi ^2d\xi \int\limits_\xi 
^\infty \frac{du}{u^2}  B\left( u \right) .
\]
The coefficient, $\alpha $, calculated numerically, is 0.28. Thus we find 
\[
I = \alpha \left( 0.45 - \frac{1}{6}\log \alpha  \right) .
\]
}
for $\alpha \ll 1$, and one finds

\[
\qquad\qquad\qquad\qquad\qquad\qquad ~ \rho ^2 = \frac{v^2}{2\pi ^2\nu _0 ^2}\alpha \left( {0.45 - 
\frac{1}{6}\log \alpha } \right) . \qquad\qquad\qquad\qquad\qquad (13^\prime)
\]

We want to apply this formula to calculate the ionization of Helium by 
$\alpha $-particles from Radium-C. Since Helium possesses $k$-shell 
electrons, and [the $k$-shell] is complete, Eq.~(13$^\prime$) will be applicable, 
always taking $H = 0.5 \times 10^{30}$. Corresponding to the ionization potential 
of 24.5 volts, we have $\nu _{0} = 6.0 \times 10^{15}$; and for 
$\alpha $-particles from Radium-C, $\nu = 1.98\times 10^{9}$. 
We therefore find $\alpha = 0.0091$; with this value one finds

\[
\rho ^2 = 0.56\times 10^{-16};
\quad
\rho = 0.75\times 10^{-8} .
\]

The number of ion pairs produced per cm by the particle is naturally 
$\pi\rho ^{_2 }n$ where $n = 2.6 \times 10^{19}$ 
represents the number of atoms per cm$^{3}$ at 15$^{\circ}$C.

This number is then 4800. Instead, experimentally, one finds 4600.

The agreement between theory and experiment is quite satisfactory; in fact, 
we observe that by chance the agreement is better than could be foreseen 
given the precision of the data.
In fact the large uncertainty, especially in the coefficient $H$, could easily 
account for a discrepancy of 20{\%} and perhaps more.

V. An additional experimental confirmation of our theory can be found in the 
treatment of deceleration of $\alpha $-particles in matter. We'll also apply 
this method to Helium for which, as was seen earlier, it is possible to find a 
reasonable value for the absorption coefficient.

First of all we want to calculate the average energy loss that an 
$\alpha $-particle undergoes passing at a distance, $b$, from an atom. 
Let $\Pi (b)$ be 
the probability that the particle passing at a distance, $b$, ionizes the 
atom, and $P(b,\nu ) d\nu $ be the probability that would exist for the 
atom to be ionized, if only frequencies in the interval, $d\nu $ were 
effective. The probability that the ionization takes place with the 
absorption of a quantum of frequency $\nu $ is then
\[
\Pi \left( b \right) \frac{P\left( {b,\nu } \right) d\nu 
}{\int\limits_0^\infty {P\left( {b,\nu } \right)d\nu } } .
\]
This ionization corresponds to an energy loss $h\nu $. The average energy 
loss of the particle is therefore
\[
\Pi \left( b \right) \frac{\int\limits_0^\infty {P\left( {b,\nu } 
\right)h\nu d\nu } }{\int\limits_0^\infty {P\left( {b,\nu } \right)d\nu } } .
\]

If $n$ represents the number of atoms per unit volume, the particle passes, 
during path~1, within a distance of $b$ to $b+db$ from 2$\pi n b db$ atoms. 
If $T$ is its kinetic energy, we have 
\begin{equation}
\frac{dT}{dx} = - 2\pi n\int\limits_0^\infty {bdb} \Pi \left( b 
\right)\frac{\int\limits_0^\infty {P\left( {b,\nu } \right)h\nu d\nu } 
}{\int\limits_0^\infty {P\left( {b,\nu } \right)d\nu } }
\end{equation}
\noindent
now we have

\[
P\left( {b,\nu } \right) = \frac{I\left( \nu \right)\alpha \left( \nu 
\right)}{h\nu } = \left\{ {{\begin{array}{*{20}c}
 {\frac{8\pi c\varepsilon^2 HZ^4}{h v^4}\frac{1}{\nu ^2}B\left( {\frac{2\pi 
\nu b}{v}} \right),\quad \nu > \nu _0 } \hfill \\
 {0,\quad\quad\quad\quad\quad\quad\quad\quad ~~ \nu < \nu _0 } \hfill \\
\end{array} }} \right.
\]

\noindent
Equation~(15) now becomes, with a simple transformation, keeping in mind Eq.~(14)
\begin{equation}
\frac{dT}{dx} = - \frac{4c\varepsilon ^2HZ^4n}{\nu _0 ^2v 
^2}\int\limits_0^\infty \xi d\xi \left\{ 1 - e^{
  - \alpha \xi \int\limits_{\xi }^\infty \frac{du}{u^2} B( u )} 
\right\} \frac{\int\limits_\xi ^\infty \frac{du}{u} 
B( u )}{\alpha \xi \int\limits_\xi ^\infty \frac{du}{u^2} 
B( u )} ~~.
\end{equation}
When $\alpha $ is very small these expressions can be evaluated approximately and one 
finds~\footnote{To calculate this integral we need to use (for $\xi < 0.4)$ the approximate expressions
\[
\int\limits_\xi ^\infty {\frac{du}{u^2}} B\left( u \right) = \frac{1}{3\xi 
^3} - 0.64,
\quad
\int\limits_\xi ^\infty {\frac{du}{u}} B\left( u \right) = \frac{1}{2\xi ^2}
- 0.64 ~.
\]
Furthermore it is convenient to divide the integral into a term from 0 to 
0.4 which can be easily calculated using the asymptotic expressions for the 
logarithmic integral. To calculate the second term from 0.4 to infinity, one 
needs only to observe that for $\xi > 0.4$ and $\alpha $ very small we can 
take
\[
1 - e{\begin{array}{*{20}c}
 { - \alpha \xi \int\limits_\xi ^\infty {\frac{du}{u^2}} B\left( u \right)} 
\hfill \\
 \hfill \\
\end{array} } = \alpha \xi \int\limits_\xi ^\infty {\frac{du}{u^2}} B\left(u\right) .
\]

One needs now to calculate numerically
\[
\int\limits_{0.4}^\infty \xi d\xi \int\limits_{\xi}^\infty 
{\frac{du}{u}} B\left(u\right) = 0.35
\]
and the result of the overall integral is
\[
0.24 - {\frac{1}{4}} log ~ \alpha 
\]
from which, applying Eq.~(15), one derives Eq.~(17).
}

\begin{equation}
\frac{dT}{dx} = - \frac{c\varepsilon ^2HZ^4n}{\nu _0 ^2v^2}\left( 
{0.96 - \log \frac{8\pi c\varepsilon ^2HZ^4}{h\nu _0 v^4}} \right)
\end{equation}

\noindent
and since $T = (1/2) mv^{2}$, we find then

\[
\qquad\qquad\qquad\qquad\qquad
\frac{dv}{dx} = - \frac{c\varepsilon ^2HZ^4n}{m\nu _0 ^2 v^3}\left( {0.96 
+ \log \frac{h\nu _0 v^4}{8\pi c\varepsilon ^2HZ^4}} \right) ~~.
\qquad\qquad\qquad\qquad (17^\prime)
\]

From this expression we can easily derive a formula which gives us the 
path-length traveled by the particles while their velocity is reduced from 
its initial value, $v_{0}$, to the final value, $v$. Explicitly one finds

\begin{equation}
x =  \frac{m\nu _0 ^2}{c\varepsilon ^2HZ^4n}\int\limits_v^{v_0 } 
{\frac{v^3dv}{0.96 + \log \frac{h\nu _0 v^4}{8\pi c\varepsilon ^2HZ^4}}} 
= 2.4\frac{m\nu _0 }{nh}\int\limits_{\frac{0.104h\nu _0 v^4}{c\varepsilon 
^2HZ^4}}^{\frac{0.104h\nu _0 v_0 ^4}{c\varepsilon ^2HZ^4}} {\frac{du}{\log u}} ~~.
\end{equation}

Naturally this formula is valid only when both limits are large with respect 
to unity, since otherwise Eq.~(17$^\prime$), from which we derived Eq.~(18) is not 
applicable, and we would then need to calculate (16) exactly also for large 
values of $\alpha $. However, we can calculate the path-length in Helium of 
$\alpha $-particles from Radium-C when their velocity is reduced by half. In 
fact, substituting in Eq.~(18) ~$m = 6.6 \times 10^{-24}$; 
$\nu _{0} = 6.0 \times 10^{15}$; $n = 2.6 \times 10^{19}$; 
$\varepsilon = 2 \times 4.77 \times 10^{-10} = 9.54 \times 10^{-10}$; 
$v_{0} = 1.98 \times 10^{9}$, we find

\[
x = 0.56\int\limits_{18}^{288} {\frac{du}{\log u}} = 32
\]

Since the \textit{range} [English word used in the original] of $\alpha $-particles of 
velocity v$_{0}$/2 is equal to about 1/8$^{th}$ of the range of particles 
with velocity v$_{0}$, we immediately deduce for the latter 37cm.

Experimentally we find, in good agreement with this theoretical value, a 
\textit{range} of approximately 33 cm.

\end{document}